\documentstyle[12pt,psfig]{article}
\title{\bf Chemical anomalies in globular cluster red giant stars and
the ``second parameter" problem}
\author{C.~Charbonnel $^1$, P.~A.~Denissenkov $^{2,3}$\\and A.~Weiss $^{3}$\\
\vspace{1cm}\\
\normalsize $^1$Laboratoire d'Astrophysique de Toulouse, CNRS UMR 5572,
France \\
\normalsize $^2$Astronomical Institute of St Petersburg University,
Russia \\
\normalsize $^3$Max-Planck-Institut f\"ur Astrophysik, Garching, Germany}

\date{\mbox{}}
\begin{document}
\maketitle
\pagestyle{empty}
%
%
\def\bull{\vrule height .9ex width .8ex depth -.1ex}
\makeatletter
\def\ps@plain{\let\@mkboth\gobbletwo
\def\@oddhead{}\def\@oddfoot{\hfil\tiny\bull\quad
``The Galactic Halo - From Globular Clusters to Field Stars'';
35$^{\mbox{\rm rd}}$ Li\`ege\ Int.\ Astroph.\ Coll., 1999\quad\bull}%
\def\@evenhead{}\let\@evenfoot\@oddfoot}
\makeatother
%
%
\def\beginrefer{\section*{References}%
\begin{quotation}\mbox{}\par}
\def\refer#1\par{{\setlength{\parindent}{-\leftmargin}\indent#1\par}}
\def\endrefer{\end{quotation}}
%
%
{\noindent\small{\bf Abstract:} 
We present the evolution of low mass and low metallicity RGB models
computed under the assumption of deep mixing between the convective
envelope and the hydrogen burning shell.
We discuss the helium enrichment of the envelope which is necessary or
allowed to achieve or to keep consistency with the observations,
and conclude on the possible connection between chemical anomalies in red 
giants and the horizontal branch morphology in globular clusters. 
%
%
\section{A possible connection between the chemical anomalies in
RGB stars and the globular cluster HB morphology}

The anomalies in elements participating in the CNO-, NeNa- and MgAl-cycles
observed in globular cluster red giants (see Sneden in this Volume for a
review) are unexplained in canonical low-mass star evolution theory and 
indicate effects beyond the standard picture. 
While some data require a primordial origin, part of the observations can 
be explained by a purely evolutionary picture which implies a non standard 
mixing process inside the low mass red giants (see Weiss et al. 1999 
and references therein). 
Depending on its depth and efficiency, this mechanism may modify the 
evolution on the red giant branch (RGB) and shape the 
horizontal branch (HB) morphology, as suggested by Langer \& Hoffman (1995) 
and discussed by Sweigart (1997). 
Basically, important helium enhacement in the envelope due to
very deep mixing would induce larger luminosity and stellar winds on the
RGB, resulting in a bluer position on the HB than in the case without 
additional mixing.

In order to test the possible connection between the chemical anomalies 
along the RGB and the HB morphology, we have computed full stellar 
evolutionary sequences which include deep mixing affecting both the 
hydrogen/helium structure of the stellar models and the distribution of 
the elements participating in the ONeNa-cycle. 
This is the first attempt to cover the problem in a consistent way, i.e., 
to treat the transport of helium (which affects the stellar structure) and 
of the other isotopes simultaneously. 

\section{Stellar models including deep mixing}
The method used for the present computations is described in details in 
Weiss et al. (1999). We consider a star of initial mass
0.8M$_{\odot}$ and of initial composition Y=0.25, Z=0.0003. 
Full evolutionary sequences are calculated with the Garching code under 
the assumption of deep mixing between the hydrogen burning shell and the
convective envelope. We do not focus on the nature of the deep mixing
mechanism. It is supposed to start at the RGB bump (Sweigart \& Mengel
1979, Charbonnel 1994, Charbonnel et al. 1998),  
and it is treated as a diffusive process with parametrized values for
the diffusive constant (D$_{\rm mix}$) and for the penetration depth 
($\Delta$X$_{\rm mix}$ which relates to the decrease in hydrogen content
and corresponds to a normalized mass coordinate $\delta$M$_{\rm mix}$ 
within the shell). Mass loss is taken into account according to
Reimers's (1975) prescription. 
The different sets of parameters (D$_{\rm mix}$, $\Delta$X$_{\rm mix}$)
used in the computations lead to various degrees of envelope helium enrichment. 
They consequently lead to higher luminosity and lower stellar mass 
M$_{\rm tip}$ at the RGB tip than in the standard case. Typically, 
M$_{\rm tip}$ can be as low as 0.56M$_{\odot}$, instead of
0.798M$_{\odot}$ in the standard case (see in Weiss et al. 1999 for a
complete description). 

These models with various degrees of mixing are then used as background
models for detailed nucleosynthesis computations in a post-processing
way, as in Denissenkov \& Weiss (1996). We show in Figure~1 the helium
enrichment of the envelope predicted in the post-processing
computations for different mixing prescriptions. 

\begin{figure}[hb]
\centerline{\psfig{figure=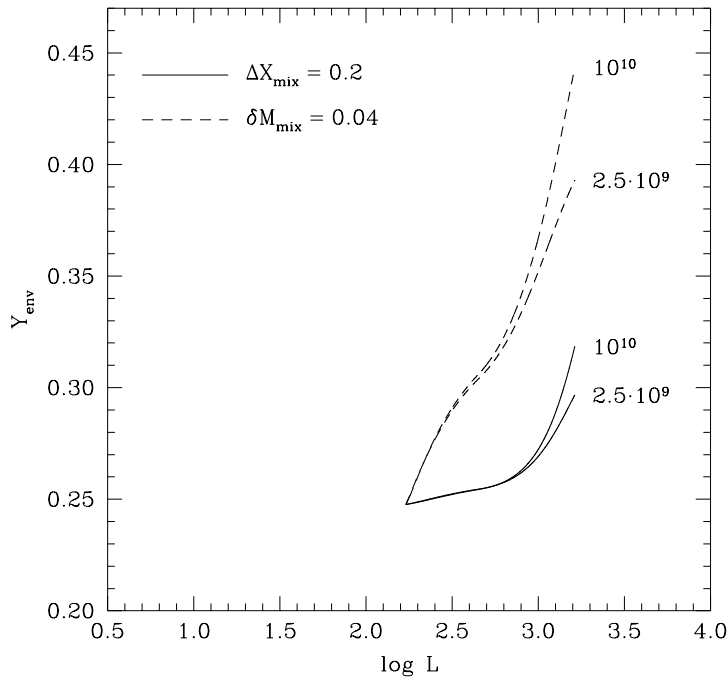,height=10cm}}
\caption{Evolution of the helium content of the envelope as a function
         of the stellar luminosity in post-processing models (initial mass
         0.8M$_{\odot}$, initial composition Y=0.25, Z=0.0003) 
         calculated with 2 different values of the diffusive constant 
         D$_{\rm mix}$ (10$^{10}$, 2.5$^9$) and of the mixing depth 
         ($\Delta$X$_{\rm mix}$, $\delta$M$_{\rm mix}$). }
\end{figure}

\section{Comparison with the global O-Na anticorrelation}
\subsection{Constraints for the mixing mechanism}

Among the chemical anomalies which have been registered over the past
two decades in red giant atmospheres, the O-Na anticorrelation is one of 
the best tracers of the mixing mechanisms that may occur deep inside the 
low mass stars before the helium flash (In contradistinction, the Mg-Al
anticorrelation first observed by Shetrone 1996 requires a combination
of the primordial and deep mixing scenarii, as discussed in Denissenkov
et al. 1998). 
This feature has been observed over a wide range of metallicity 
(-2.5$<$[Fe/H]$<$-1), in monometallic as well as in multi-metallicity 
globular clusters (see Figure 2).  
Its morphology can be explained straightforwardly within the deep mixing 
scenario, on which it gives an important insight. 
Basically, the extension of the anticorrelation along the horizontal axis
depends essentially on the mixing rate times the mixing time, while 
its extension along the vertical axis traces the depth reached by the 
mixing. 
Indeed, on approaching the hydrogen burning shell the Na abundance 
experiences two distinct rises which result from proton-captures in the 
NeNa-cycle by $^{22}$Ne and $^{20}$Ne respectively 
(Denissenkov \& Denissenkova 1990, Langer et al. 1993). 
The vertical extension of the anticorrelation indicates that the
extra-mixing, whatever its nature, does not penetrate the second step of
the Na abundance profile. 
The only cluster in which some giants exhibit surface enrichment
produced from both $^{22}$Ne and $^{20}$Ne is $\omega$Cen (Norris \& Da
Costa 1995; see Figure 2); most of those stars with extremely high [Na/Fe] 
(up to 1 dex, indicating that the second rise of Na is reached by the deep 
mixing process) belong to the metal-rich population of $\omega$Cen. 
This very peculiar cluster also owns one of the bluest horizontal branches 
(Whitney et al. 1994). 

\begin{figure}[hb]
\centerline{\psfig{figure=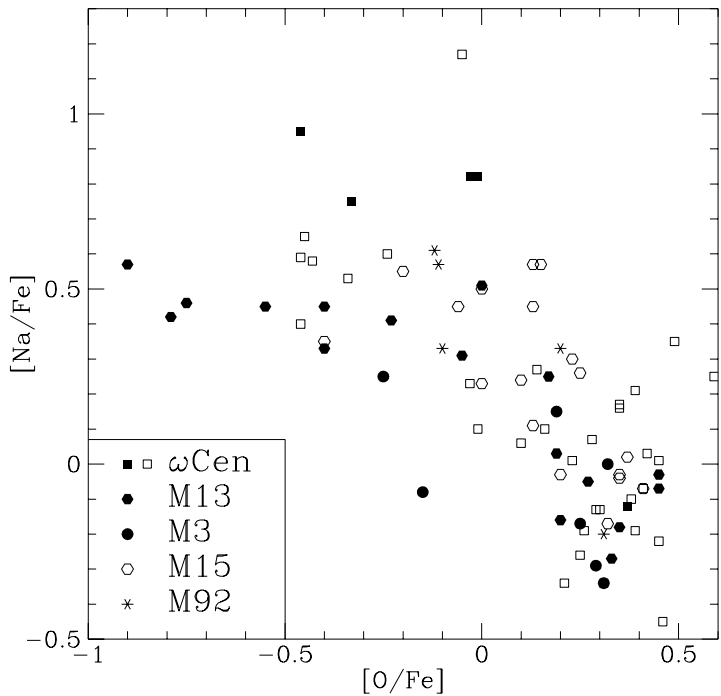,height=10cm}}
\caption{The global O-Na anticorrelation in several globular clusters
(the black and white squares correspond to stars of $\omega$Cen with
[Fe/H] respectively higher and lower than $-1.0$). The data are from
Kraft et al. (1992, 1997), Norris \& Da Costa (1995), Shetrone (1996), 
Sneden et al. (1997).}
\end{figure}

\subsection{Predictions vs observations}
In Figure 3 we compare the O-Na anticorrelation observed in three
globular clusters with our predictions from models with different degrees 
of mixing.  
The models which undergo an extra-mixing leading to important helium
enrichment of the envelope also exhibit Na overabundances which are never 
observed, except maybe in the most metal-rich giants of 
$\omega$Cen. In this cluster, the stars above [Na/Fe]=0.6 could be 
explained by shallow and long-lasting mixing; however [O/Fe]=0 is difficult to
keep unless a primordial effect is added. Note that a metallicity effect 
(not considered here) may also play a role.

For what concerns the second parameter globular cluster M13 which presents 
the most extreme oxygen underabundances ([O/Fe]$\leq$-0.45), no giant
shows [Na/Fe] higher than $\sim$0.5. This is in contradiction with what is
expected in the models with high helium enrichment. 
We can conclude that the global anticorrelation rules out the
possibility to get more than $\delta$Y$\simeq$0.06 increase of the
helium abundance in globular cluster red giants. 

  \begin{figure}[hb]
  \centerline{\psfig{figure=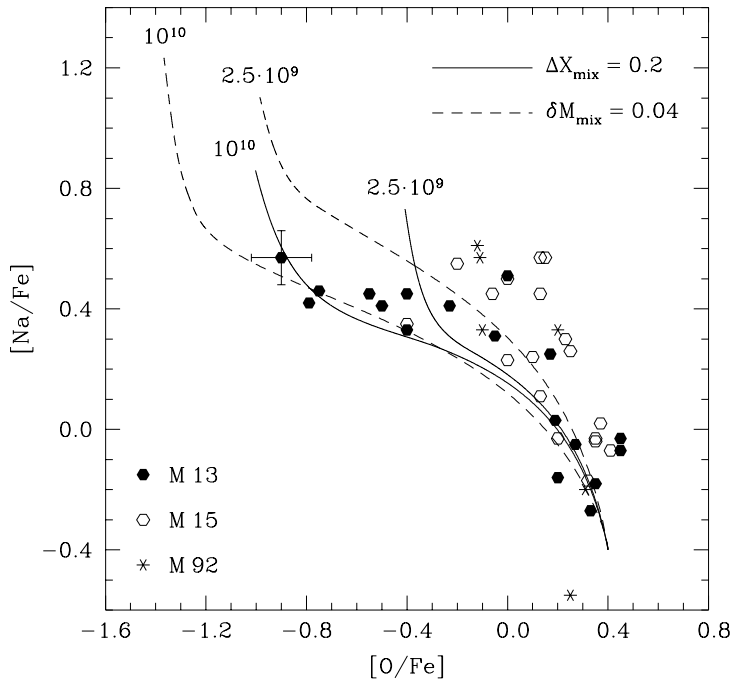,height=10cm}}
  \caption{The global O-Na anticorrelation in the globular clusters
           M13, M15 and M92, and the theoretical curves obtained under 
           the same assumptions than in Figure 1.}
  \end{figure}

\section{Conclusions}
The global O-Na anticorrelation gives a severe constraint to the deep
mixing in RGB stars. This feature can be explained by mixing which does 
not reach the second rise in the Na abundance profile, and thus does not 
lead to significant helium enrichment of the envelope. A maximum
increase of the envelope helium abundance $\Delta$Y of about 0.06 only is
expected. This value is much lower than ascribed by Sweigart (1997) to
get a significant influence on the stellar evolution on the RGB and on 
the subsequent position along the HB.  
Models with very deep mixing and such strong helium increase predict
anomalies of sodium and oxygen which are much larger than the observed
ones. 

%
%
\section*{Acknowledgements}
This study was done while C.C. and P.A.D. visited the
Max-Planck-Institute f\"ur Astrophysics in Garching. 
They wish to express their gratitude to the staff for their hospitality 
and support. 
%
%
 
\beginrefer
\refer Charbonnel C., 1994, A\&A 282, 811 

\refer Charbonnel C., Brown J., Wallerstein  G., 1998, A\&A 332, 204 

\refer Denissenkov P.A., Denissenkova S.N., 1990, SvA Lett. 16, 275

\refer Denissenkov P.A., Weiss A, 1996, A\&A 308, 773

\refer Denissenkov P.A., Da Costa G.S., Norris J.E., Weiss A., 1998, A\&A 333, 
926

\refer Kraft R.P., Sneden C., Langer G.E., Prosser C.F., 1992, AJ 104,
645

\refer Kraft R.P., Sneden C., Smith G.H., Shetrone M.D., Langer G.E.,
Pilachowski C.A., 1997, AJ 113, 279

\refer Langer G.E., Hoffman R.D., 1995, PASP 107, 1177

\refer Langer G.E., Hoffman R.D., Sneden C., 1993, PASP 105, 301

\refer Norris J.E., Da Costa G.S., 1995, ApJ 447, 680

\refer Reimers D., 1975, Mem. Soc. Roy. Sci. Li\`ege 8, 369

\refer Shetrone M.D., 1996, AJ 112, 1517

\refer Sneden C., Kraft R.P., Shetrone M.D., Smith G.H., Langer G.E.,
Prosser C.F., 1997, AJ 114, 1964

\refer Sweigart A.W., 1997, ApJ 474, L23

\refer Sweigart A.W., Mengel K. G., 1979, ApJ 229, 624

\refer Weiss A., Denissenkov P.A., Charbonnel C., 1999, A\&A, in
preparation

\refer Whitney J.O., O'Connel R.W., Wood R.T., 1994, AJ 108, 1350

\endrefer
\end{document}